\documentclass[twocolumn, twocolappendix]{aastex631}
\usepackage{natbib}
\usepackage{wrapfig}
\usepackage{amsmath}
\usepackage{hyperref}
\usepackage{enumerate}
\begin{document}

\title{Improving Photometric Redshift Estimates with Training Sample Augmentation}

\correspondingauthor{Irene Moskowitz}
\email{iwm15@physics.rutgers.edu}

\author[0000-0002-2206-8589]{Irene Moskowitz}
\affiliation{Department of Physics and Astronomy, Rutgers, The State University of New Jersey, Piscataway, NJ 08854, USA}

\author[0000-0003-1530-8713]{Eric Gawiser}
\affiliation{Department of Physics and Astronomy, Rutgers, The State University of New Jersey, Piscataway, NJ 08854, USA}

\author[0000-0002-2495-3514]{John Franklin Crenshaw}
\affiliation{Department of Physics, University of Washington, Seattle, WA 98195, USA}
\affiliation{DIRAC Institute, University of Washington, Seattle, WA 98195, USA}

\author[0000-0001-8085-5890]{Brett H.~Andrews}
\affiliation{Department of Physics and Astronomy and PITT PACC, University of Pittsburgh, Pittsburgh, PA 15260, USA}

\author[0000-0002-8676-1622]{Alex I. Malz}
\affiliation{McWilliams Center for Cosmology, Department of Physics, Carnegie Mellon University, Pittsburgh, PA, USA}

\author[0000-0002-5091-0470]{Samuel Schmidt}
\affiliation{Department of Physics and Astronomy, University of California, Davis, CA, 95616, USA}

\author{The LSST Dark Energy Science Collaboration}

\submitjournal{ApJ Letters}

\begin{abstract}
Large imaging surveys will rely on photometric redshifts (photo-$z$'s), which are typically estimated through machine learning methods.  Currently planned spectroscopic surveys will not be deep enough to produce a representative training sample for LSST, so we seek methods to improve the photo-$z$ estimates that arise from non-representative training samples. Spectroscopic training samples for photo-$z$'s are biased towards redder, brighter galaxies, which also tend to be at lower redshift than the typical galaxy observed by LSST, leading to poor photo-$z$ estimates with outlier fractions nearly 4 times larger than for a representative training sample. In this paper, we apply the concept of training sample augmentation, where we augment simulated non-representative training samples with simulated galaxies possessing otherwise unrepresented features. When we select simulated galaxies with (\textit{g}-\textit{z}) color, \textit{i}-band magnitude and redshift outside the range of the original training sample, we are able to reduce the outlier fraction of the photo-$z$ estimates for simulated LSST data by nearly $50\%$ and the normalized median absolute deviation (NMAD) by $56\%$. When compared to a fully representative training sample, augmentation can recover nearly $70\%$ of the degradation in the outlier fraction and $80\%$ of the degradation in NMAD. Training sample augmentation is a simple and effective way to improve training samples for photo-$z$'s without requiring additional spectroscopic samples.  
    
\end{abstract}

\section{Introduction}
Understanding the nature of dark energy is a major open question in cosmology. Stage-IV dark energy experiments, such as the Vera C. Rubin Observatory's Legacy Survey of Space and Time (LSST, \citealt{lsst}), Euclid \citep{euclid} and Roman \citep{roman}, are scheduled to come online in the coming years. 

Imaging surveys will need to obtain redshifts to galaxies, but there will be too many for spectroscopic redshifts to be feasible. LSST alone is expected to observe billions of galaxies and will therefore rely on photometric redshifts (photo-$z$'s). Photo-$z$'s can be estimated through machine learning algorithms, which learn to associate photometric quantities, such as colors and magnitudes, with a redshift estimate.

Previous Stage-III dark energy surveys have also used machine learning for estimating photometric redshifts. The Hyper Suprime-Cam Subaru Strategic Program (HSC-SSP, \citealt{hsc_overview}) used DNNz and DEMPz \citep{dempz} for the Year 3 cosmology results \citep{hsc_photoz, hsc_y3_results1, hsc_y3_results2}. Both DNNz and DEMPz are conditional density estimators. The Dark Energy Survey (DES, \citealt{DES}) has used a self-organizing map (SOMPZ, \citealt{des_y3_photoz}) for estimating photo-$z$'s. The Kilo-Degree Survey (KiDS, \citealt{kids}) has also used self-organizing maps for photo-$z$ estimation \citep{kids1000_photoz}.

Machine learning methods require a training sample of galaxies with both photometry and spectroscopic redshifts, and it is well known that machine learning methods trained on non-representative training data perform worse than when trained on representative training sets (see e.g., \citealt{beck2017} for a general evaluation of photo-$z$ quality when non-representative training samples are used). \citet{Stylianou2022} also demonstrates the effect of some simplistic forms of training sample incompleteness on specific machine learning methods. However, existing spectroscopic samples are biased towards brighter, redder galaxies than LSST will observe in general, and these also tend to be at lower redshift than the typical LSST galaxy. This means that training samples for photo-$z$ estimation will not be representative of LSST data, leading to poor photo-$z$ estimation for galaxies with photometry not represented in the training sample. The Dark Energy Spectroscopic Instrument (DESI, \citealt{desi}), along with spectroscopic redshifts from Euclid, Roman and 4MOST \citep{4most}, will alleviate this issue to an extent, but the DESI survey will not be as deep as LSST; additional spectroscopic redshifts from DESI can not solve the problem alone. We will need methods to improve the redshift estimation that do not involve obtaining more spectroscopic redshifts.

One method for improving training samples without obtaining more spectroscopic redshifts is through data augmentation, which is the process of modifying a training sample in some way to increase the generality of a machine learning model \citep{augmentation}. Data augmentation can be done by transforming existing training sample data in some way, such as through rotations or deformations in the case of image recognition \citep{image_augmentation}, or by generating synthetic data for the training sample \citep{synthetic_augmentation}. \citet{Broussard_2021} used this synthetic data generation method for augmentation to estimate photo-$z$'s. 

In this paper, we investigate a slightly different method of augmenting the training sample by adding galaxies from simulated catalogs to our training sample. By selecting simulated galaxies with photometry and/or redshifts not otherwise represented in the training sample, this training sample augmentation can expand the range of feature space capable of producing good photo-$z$ estimates, provided the simulated catalog used for augmentation has reasonable colors. If the simulated catalog is too unrealistic, this will only create confusion in our model. 

Section 2 describes our simulated data, including our stand in for real LSST data and the simulated catalog used for augmenting the training sample. Section 3 describes our methodology, including how a realistically non-representative training sample is created, how we estimate photo-$z$'s, and the process for augmenting the training sample. Section 4 discusses our results, and section 5 concludes.

\section{Simulated Data}
\subsection{DC2}
The LSST Dark Energy Science Collaboration (DESC) Data Challenge 2 catalog (DC2;\citealt{DC2}) is a 300 deg$^2$ area of simulated LSST observations. The base input for DC2 is the CosmoDC2 galaxy catalog \citep{cosmoDC2}, which is derived from the Outer Rim N-body simulation \citep{outer_rim}. Galaxies were assigned to halos using UniverseMachine \citep{universemachine} and GalSampler \citep{galsampler}. Complete galaxy properties are generated with Galacticus \citep{galacticus}. Galaxy spectra are constructed from stellar population spectra computed with \textsc{fsps} \citep{fsps}.

To generate the DC2 catalog, stars, supernovae, strong lenses and AGN are added to the CosmoDC2 catalog. The object catalogs are passed as inputs to the image simulation software imSim \footnote{\href{https://github.com/LSSTDESC/Imsim}{https://github.com/LSSTDESC/Imsim}} to generate LSST-like images, which are then processed by the LSST Science Pipelines.

From the DC2 catalog, we select objects with magnitude $i> 17$. We require S/N $> 6$ in the \textit{i}-band, as well as S/N $>3$ in at least one other band. To minimize contamination from stars, we select only extended objects. This extendedness cut does not entirely eliminate stars, but the stellar contamination is low, and the training sample selection process (described in section \ref{sec:non_rep_training}) ends up placing all stars in the application sample. 

The base, unaugmented training sample and the application sample of galaxies, for which we estimate redshifts, are formed from our selected DC2 objects. In this work, DC2 is a stand-in for real data. We use the term `application sample' to refer to the DC2 stand-in for what would be unlabeled LSST data. While we do have true redshifts for this sample, \textbf{and it functions as a testing set in this analysis}, we keep the terminology to identify our application sample with eventual LSST data.

\subsection{Buzzard}
The Buzzard simulation \citep{buzzard} was built from the L-GADGET2 dark matter simulations \citep{gadget2}. Galaxies are assigned to halos using \textsc{AddGals} \citep{add_gals} using the abundance matching technique. Galaxy SEDs were assigned to match the measured SED-luminosity-density relationship in SDSS data. 

We use the Buzzard catalog selected for the DESC Tomographic Challenge \citep{tomo_challenge}. Details on selection cuts, post-processing and uncertainty generation can be found in that paper.

The Buzzard method of assigning spectra is completely independent from \textsc{fsps}, so Buzzard SEDs should be sufficiently different from DC2 SEDs to simulate adding simulated galaxies to training samples of real galaxies. In this work, we use the Buzzard catalog as a simulated catalog with which to augment the DC2 training sample.

\section{Methodology}

\subsection{Non-representative Training Sample}\label{sec:non_rep_training}

\begin{figure*}
    \plotone{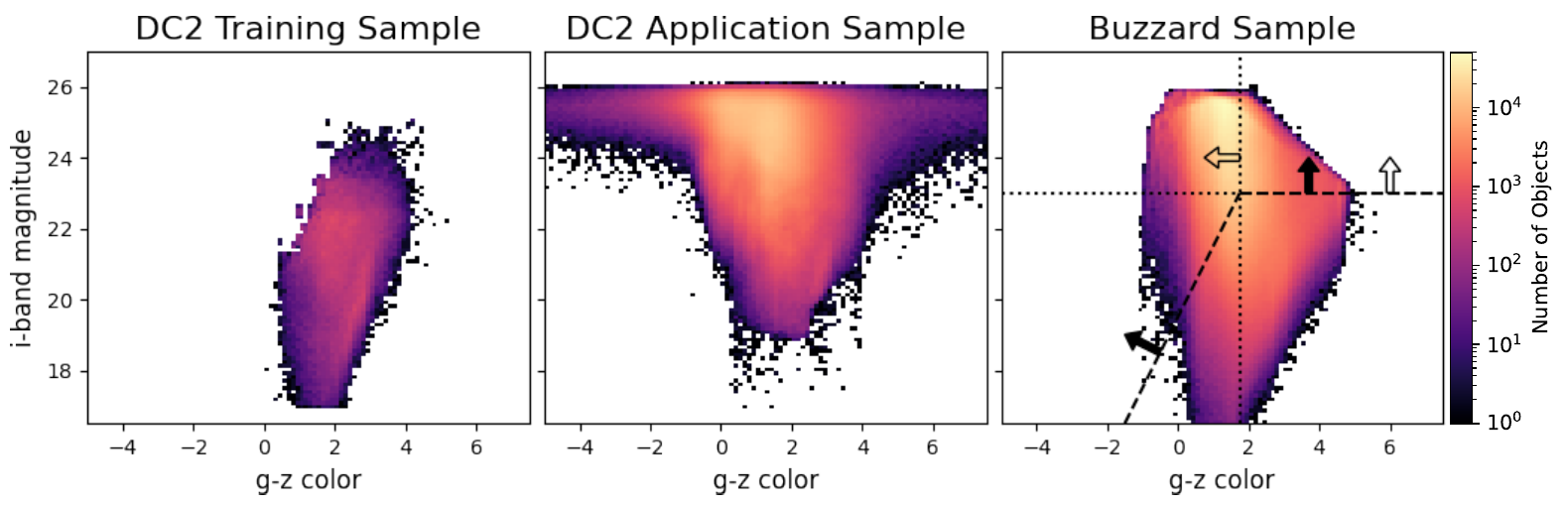}
    \caption{The results of partitioning our DC2 catalog into training (left) and application (center) samples. The training sample is redder and brighter than the bulk of the application sample. The right panel shows the Buzzard sample used for augmenting the training sample. The horizontal dotted line shows the \textit{i}-band selection criterion, while the vertical dotted line shows the (\textit{g}-\textit{z}) color criterion. The dashed line indicates the section criterion for color+magnitude augmentation, which generally matches the shape of the DC2 training sample in the left panel. Open arrows indicate which regions of color-magnitude space are used for single feature augmentation, while solid arrows indicate regions used for color+magnitude augmentation See section 3.3 for more details on augmentation criteria.}
    \label{fig:nonrep_training}
\end{figure*}

Existing spectroscopic galaxy samples are brighter and redder than expected LSST observations, and also tend to be at lower redshifts. To partition our DC2 catalog into a realistically non-representative training sample and application sample, we use the GridSelection degrader in the DESC RAIL\footnote{\href{https://github.com/LSSTDESC/rail}{https://github.com/LSSTDESC/rail}} software \citep{rail}. We briefly summarize the GridSelection degrader below. A more detailed discussion can be found in \citet{Moskowitz_2023}.

The GridSelection degrader is based on the second data release of the Hyper Suprime Cam Subaru Strategic Program \citep{hsc_dr2}. Galaxies with similar photometry to early LSST observations are selected from the HSC Wide catalog; some of these galaxies have photometry only, and some have matched spectroscopic redshifts. The range in \textit{i}-band magnitude and (\textit{g-z}) color are divided into 100x100 pixels. Within each pixel, a ratio of the number of galaxies with spectroscopic redshifts to the total number of galaxies is computed, along with the 99th percentile in spectroscopic redshift, denoted $z_{max}$. The GridSelection degrader divides our DC2 galaxies into the same set of pixels in \textit{i} vs (\textit{g}-\textit{z}) and automatically assigns DC2 objects with $z_{true} > z_{max}$ to the application sample. From the remaining DC2 objects, the GridSelection degrader randomly selects objects for the training sample such that the ratio of DC2 training objects to total objects in a pixel matches the ratio from HSC. 

\begin{figure}
    \plotone{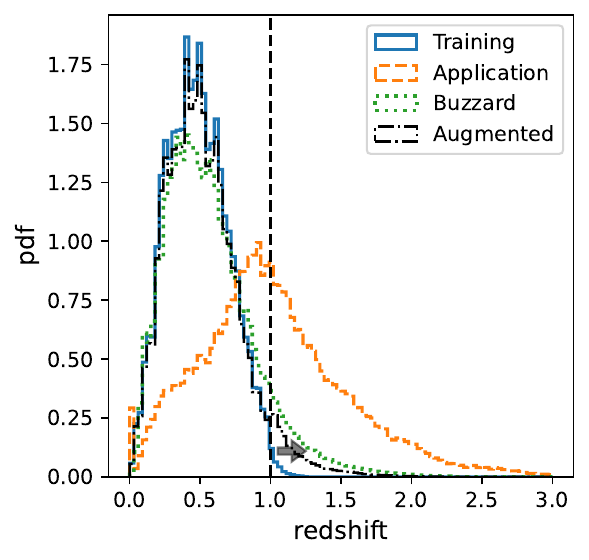}
    \caption{Normalized redshift distributions of the DC2 training sample (blue solid line), DC2 application sample (orange dashed line), and Buzzard sample (green dotted line). The DC2 training sample is biased to lower redshifts than the application sample. The vertical dashed line indicates the selection criterion for redshift augmentation, with the arrow indicating the region of redshift space used for augmentation. The black dot-dashed line shows the redshift distribution of the best performing, post augmentation training sample shown in the top right panel of Figure \ref{fig:best_results}.}
    \label{fig:nonrep_redshifts}
\end{figure}

After partitioning the full DC2 sample into training and application samples, the training sample contains 186,837 galaxies, while the application sample contains 5,520,458 objects. The left and center panels of Figure \ref{fig:nonrep_training} show the resulting DC2 training and application samples, where it is clear that the training sample is redder and brighter than the majority of the application sample. Figure \ref{fig:nonrep_redshifts} shows the (normalized) redshift distributions of both samples, as well as the distributions of the Buzzard sample and the best performing augmentation choice. The training sample is biased towards lower redshifts than the application sample as a whole. The right panel of Figure \ref{fig:nonrep_training} shows the Buzzard sample that will be used for augmentation. See section 3.3 for more details.

\subsection{Photo-z estimation}
\citet{schmidt2020} tested twelve photo-$z$ estimation codes, albeit using representative training data, and recommended FlexZBoost (\citealt{fzboost2017, fzboost2020}) as an appropriate estimator. Therefore, to estimate photo-$z$'s, we use FlexZBoost  as implemented in RAIL. FlexZBoost is a non-parametric conditional density estimator for redshifts. It takes as inputs the magnitudes and errors in each of the bands \textit{ugrizy} and outputs a photo-z pdf for each object in the application sample.

To evaluate the quality of a set of photo-z estimates, we use the outlier fraction, catastrophic outlier fraction, normalized median absolute deviation (NMAD) and bias. We define an outlier as $|z_{true} - z_{phot}| / (1+z_{true}) > 0.15$, while a catastrophic outlier is defined as $|z_{true} - z_{phot}| > 1.0$. The NMAD is given by:
\begin{equation}
    1.4826 \times \textrm{Med}\left[\lvert\frac{\Delta z}{1+z_{true}} - \textrm
    {Med}\left(\frac{\Delta z}{1+z_{true}}\right)\rvert\right],
\end{equation}
where the bias is given by Median$(\Delta z / (1+z_{true}))$. Although the pdf contains a wealth of useful information that can be used to quantify photo-$z$ quality, such as the 3$\sigma$ outlier fraction (see e.g. \citealt{jones2023}), cosmological analyses typically involve assigning galaxies to tomographic redshift bins. Since galaxies can only be assigned to one redshift bin, little information is lost by compressing the pdf into a single photo-$z$ point estimate used to assign the galaxy to a bin. We use the mean of the pdf as a point estimate for each photo-$z$.

\begin{figure}
    \plotone{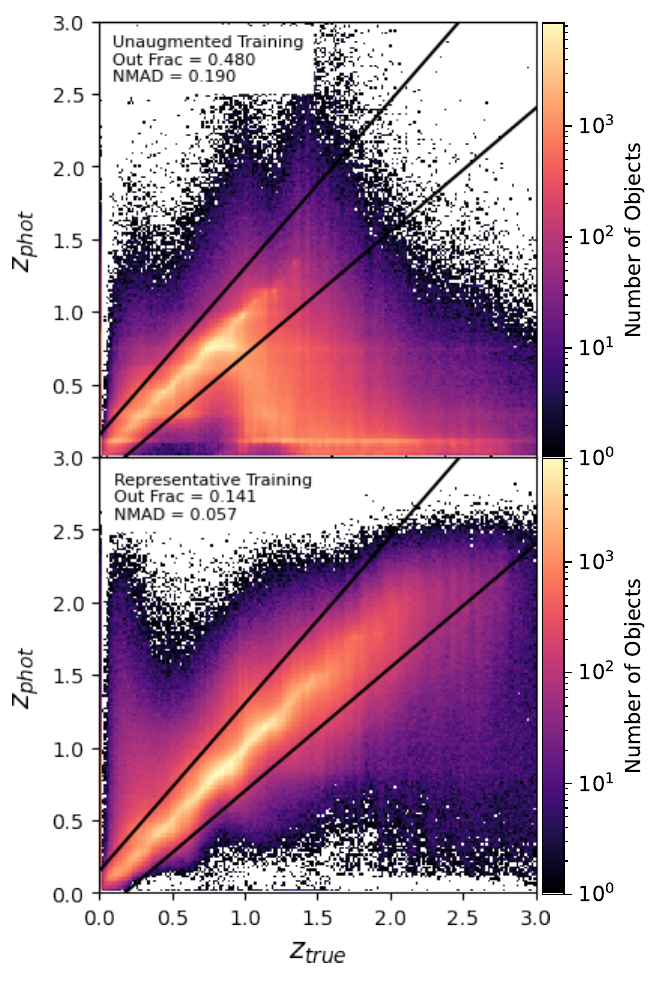}
    \caption{\textit{Top}: Photo-$z$'s estimated from the realistic, non-representative DC2 training sample shown in Figure \ref{fig:nonrep_training}. Solid black lines indicate the boundary for outliers. \textit{Bottom}: Photo-$z$'s estimated from a fully representative training sample drawn randomly from the DC2 application sample in Figure \ref{fig:nonrep_training}.}
    \label{fig:base_photoz}
\end{figure}

The photo-$z$'s estimated from the base, non-representative DC2 training sample are shown in the top panel of Figure \ref{fig:base_photoz}. The outlier fraction is quite high at nearly 50\%. In particular, the majority of galaxies with $z_{true} \gtrsim 1.0$ have outlier $z_{phot}$ estimates. This is due to the fact that our DC2 training sample has very few objects with $z > 1.0$. For comparison, the bottom panel of Figure \ref{fig:base_photoz} shows the results for a fully representative DC2 training sample, which obtains an outlier fraction of 0.14. This represents the best we can expect to do using FlexZBoost. As shown in Figure \ref{fig:base_photoz}, the unaugmented, non-representative training sample produces much worse photo-$z$'s than the representative training sample, particularly at $z_{true}> 1.0$.

\subsection{Augmentation}
We augment the training sample by adding 10,000 Buzzard galaxies according to a set of criteria, taking care to only use knowledge about the application sample that would be available for real data. The simplest criterion for augmentation is to select Buzzard galaxies with higher redshifts than those present in the DC2 training sample. We refer to this as redshift augmentation, and make the selection $z_{buzzard} > 1.0$. This region is indicated by the vertical dashed line and arrow in Figure \ref{fig:nonrep_redshifts}.

Since DC2 application galaxies are also dimmer and bluer than the training sample, we also choose magnitude and color selection criteria, which we call magnitude augmentation and color augmentation, respectively. For magnitude augmentation we make the selection $i_{buzzard} > 23$, and for color augmentation we choose $(g-z)_{buzzard} < 1.75$. These boundaries were chosen to match where the magnitude and (\textit{g}-\textit{z}) color distributions in the training sample start to decline. They are indicated by the dotted lines and open arrows in the right panel of Figure \ref{fig:nonrep_training}. We test augmentation with each of the features individually, as well as in combination with each other. 

\subsection{Photometry Shifts}
Since DC2 and Buzzard rely on different methods for determining galaxy SEDs, the colors as a function of redshift are different between the two simulations. If this augmentation method was used for real data, it would be advantageous to shift the simulated photometry to look like the real photometry in the application sample. Therefore, we also attempt to match the Buzzard photometry to the DC2 application sample. This modifies the color-redshift relationship in Buzzard to potentially more closely resemble the color-redshift relationship of DC2. Since we do not use the true redshifts of the application sample, this represents something we could do with real data.

\subsubsection{Magnitude Shifts}
The simplest way to transform Buzzard colors is to apply a single shift to the Buzzard magnitudes to make their median match the median of the DC2 application sample magnitudes in each band. We will refer to this sample as the `magnitude shifted Buzzard' sample.

In addition to the medians, we can also rescale the NMADs to match in  each band. This is a proxy for matching the first and second moments of the photometry distributions. To shift the NMADS, we apply the following transformation:
\begin{equation}
    mag_{j,new} = \frac{NMAD_{j,DC2}}{NMAD_j}\times[mag_j -med_j] + med_j
\end{equation}
where $NMAD_j$ refers to the NMAD in band $j$, $med_j$ is the median magnitude in band $j$ and all quantities are for Buzzard unless indicated by the DC2 subscript. We will refer to this sample as the `NMAD shifted Buzzard' sample.

\subsubsection{Normalizing Flows}
The simple shift method is able to match the medians and NMADs of the Buzzard and DC2 color distributions, but not the shapes of the distributions. To attempt to more fully match the color distributions, we use normalizing flows to produce a catalog of DC2-like photometry with Buzzard-like redshifts.

We use the PZFlow\footnote{\href{https://jfcrenshaw.github.io/pzflow/}{https://jfcrenshaw.github.io/pzflow/}}  package \citep{pzflow} as implemented in RAIL for training the normalizing flows. We train two flows: one on DC2 photometry, and one on Buzzard photometry. The DC2 flow learns the probability distribution function of the DC2 photometry, p(photometry), while the Buzzard flow is a conditional flow that learns the probability density function of the redshift given the photometry, p($z|$photometry). The features used for training are \textit{i}-band magnitudes and (\textit{u-g}), (\textit{g-r}), (\textit{r-i}), (\textit{i-z}) and (\textit{z-y}) colors. We train 100 epochs for the DC2 flow, and 150 epochs for the Buzzard flow.

Once the flows are trained, we sample from the DC2 flow to make a new catalog of galaxies with DC2-like photometry. We then sample from the Buzzard flow, using the new DC2-like photometry as conditions, to generate Buzzard-like redshifts for our DC2-like photometry. Finally, we use the RAIL LSSTErrorModel degrader to generate LSST-like errors on the magnitudes. This set of DC2-like photometry and Buzzard-like redshifts constitutes our flowed catalog from which we draw galaxies for augmentation. We will refer to this sample as the ``flowed Buzzard" sample.

\section{Results}
\begin{figure*}
    \plotone{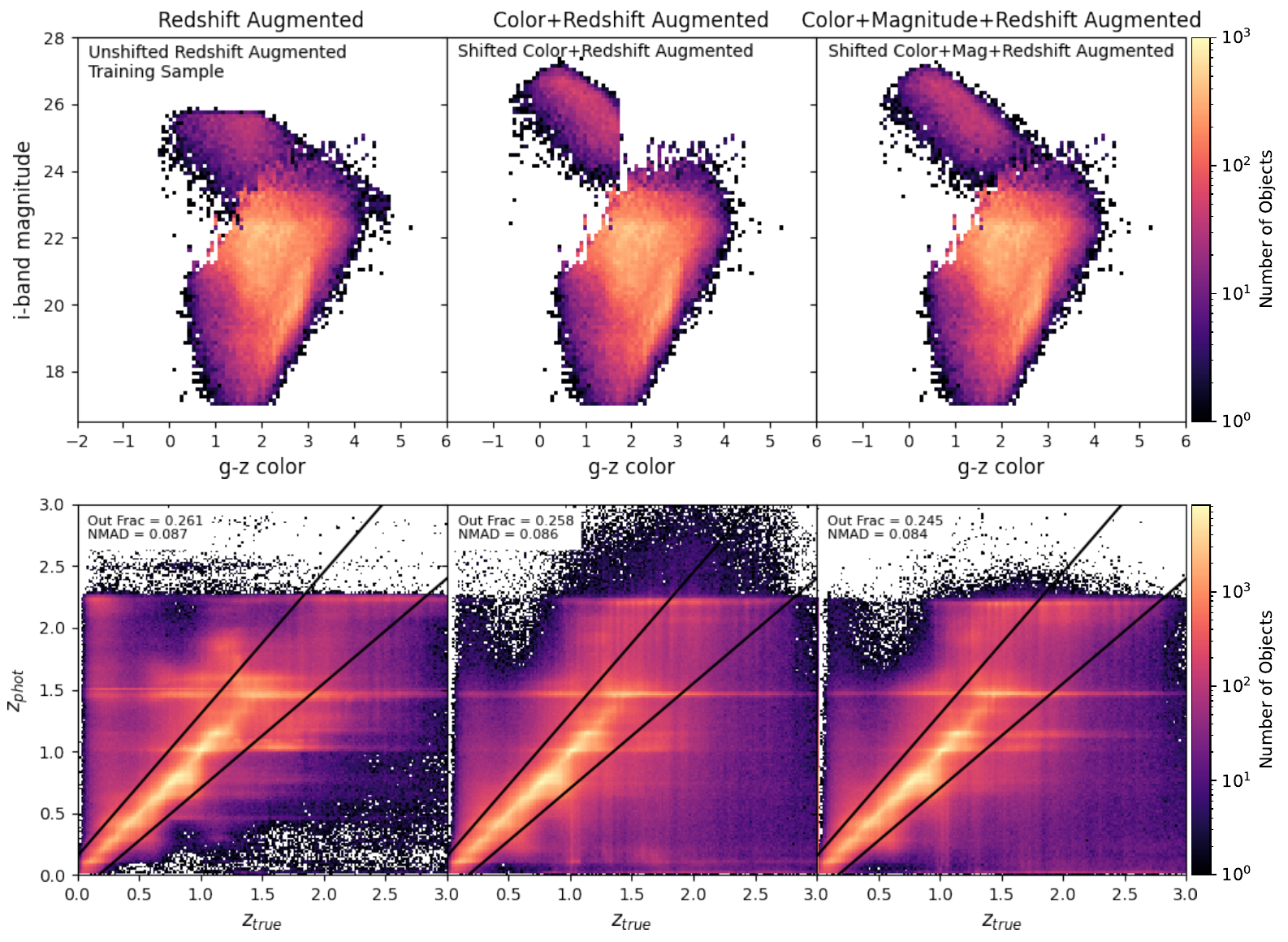}
    \caption{The best case photo-$z$ estimation for single, double and triple feature augmentation (bottom row), with the corresponding augmented training samples (top row). \textit{Left:} When only a single feature is used for augmentation, redshift augmentation produces the best photo-$z$ estimates. The training sample was augmented with the unshifted Buzzard catalog, which in this case produced better photo-$z$ statistics than the magnitude shifted or flowed Buzzard catalog. \textit{Center:} When a double feature combination is used for selecting galaxies for augmentation, the combination of color+redshift produces the best results. The best results for this case came from using the magnitude shifted Buzzard catalog.   \textit{Right:} The best photo-$z$ estimates were produced when all three features are combined for selected galaxies for augmentation. The best results for this case came from using the magnitude shifted Buzzard catalog.}  
    \label{fig:best_results}
\end{figure*}

Table \ref{tab:summary} summarizes the outlier fractions and NMADS achieved for each combination of augmentation features and each Buzzard sample (unshifted, magnitude shifted and flowed), as well as those achieved for the unaugmented training sample and a fully representative training sample.
Every kind of augmentation we tested improved the outlier fraction and NMAD of the resulting photo-$z$'s. Augmentations involving redshift selections performed better than those without. The magnitude shifted Buzzard sample generally produced better results than the unshifted or flowed Buzzard samples; however, in the case of selecting galaxies for augmentation using only a single features, the unshifted Buzzard catalog produced the best results. 

Adding the NMAD shift to the magnitude shift did not hurt, but showed no improvement over the simple magnitude shift, so we only show results for the magnitude shifted sample. We also tried combining the magnitude shifted Buzzard sample with the normalizing flow method, but the standard color flow worked better. 

Since FlexZBoost also takes in photometric errors, we tested the effect of changing the errors. We tested multiplying the errors by factors of 0.1, 2.0 and $2.0 \times (1+z)$. Variations in the outlier fractions and NMADs were smaller than the variations between augmentation cases.

Each method of augmentation produces an outlier fraction and NMAD for the resulting photo-$z$'s. In addition to the raw metrics, we also report the ratio of the augmented results to the unaugmented results. Since the best case scenario is the fully representative training case, not an outlier fraction/NMAD of 0, we also report a percent recovery towards the outlier fraction/NMAD achieved in the representative case. The percent recovery is calculated as $(X_{unrep} - X_{aug})/(X_{unrep}-X_{rep})$, where $X$ is either the outlier fraction or NMAD, the subscript `aug' refers to the statistic for the augmentated training sample, the subscript `unaug' refers to the unaugmented training sample, and the subscript `rep' refers to the representative training sample.

The following subsections discuss the results for single feature, double feature and triple feature augmentations. Table \ref{tab:summary} summarizes the results. 

\subsection{Augmentation with Individual Features}
When augmenting with a single feature, we choose Buzzard galaxies with either $z_{true} > 1.0$, $i$-mag $> 23$ or $(g-z) < 1.75$. When using a single feature for augmentation, the base, unshifted Buzzard catalog produced the lowest outlier fractions and NMADs.

Of the three features, redshift augmentation produces the best results when a single feature is used for augmentation, while color performs the worst. The redshift augmented training sample, using the unshifted Buzzard catalog, and the resulting photo-$z$ estimates for the redshift augmentation are shown in the left panel of Figure \ref{fig:best_results}. Results for color and magnitude augmentation are, as well as for the shifted and flowed Buzzard catalogs are listed in Table \ref{tab:summary}.


\subsection{Augmentation with Double Feature Combinations}

\begin{table*}
    \centering
    \begin{tabular}{c|ccc|ccc|ccc} 
 \textbf{Unaugmented} & Outlier& NMAD & Bias &&&& \\
  \textbf{Samples} & Fraction & & &&&& \\
    \hline
    Representative& 0.141(0.014)& 0.057 & 0.0001&&&& \\ 
    Non-Representative& 0.480(0.21)& 0.190 & -0.12&&&& \\
    \hline
 \textbf{Augmented}& \multicolumn{3}{|c|}{Unshifted Buzzard} & \multicolumn{3}{|c|}{Magnitude Shifted Buzzard}& \multicolumn{3}{|c}{Flowed Buzzard}\\
 \textbf{Samples} &&&&&&& \\
    \hline
         &  Outlier &  NMAD & Bias &  Outlier &  NMAD & Bias&  Outlier & NMAD & Bias\\ 
         & Fraction& & &  Fraction& & &  Fraction& & \\ \hline
         $z$ ($z_{buz} > 1.0$)&  0.261(0.040)&  0.087 & -0.004& 0.263(0.046)&  0.088& -0.014 & 0.292(0.045)& 0.094 & 0.003\\  
         Mag ($i_{buz} > 23$)&  0.318(0.12)&  0.097 &  -0.030&0.407(0.17)&  0.138& -0.069& 0.327(0.045)& 0.107 &-0.039\\  
         Col ($(g-z)_{buz} < 1.75$)&  0.324(0.12)&  0.099 & -0.031& 0.401(0.17)&  0.134& -0.066& 0.319(0.069)& 0.102 &-0.035\\  
         Mag+$z$&  0.268(0.037)&  0.090 & -0.001 &  0.259(0.047)&  0.086& -0.015&  0.293(0.047)& 0.096 & 0.001\\ 
         Col+$z$&  0.271(0.037)&  0.090 & -0.005& 0.258(0.045)&  0.086& -0.016& 0.286(0.046)& 0.093 &0.0004\\ 
         Col+Mag&  0.311(0.12)&  0.093 & -0.028&  0.400(0.16)&  0.134& -0.067& 0.327(0.066)& 0.107&-0.040\\  
         \textbf{Col+Mag+$z$}&  0.268(0.039)&  0.089 & -0.002& \textbf{0.245(0.037)}&  \textbf{0.084}& \textbf{-0.014}& 0.284(0.043)& 0.092 &0.002\\ 
         \hline
    \end{tabular}

    \begin{tabular}{c|c|ccccc}
        
    \end{tabular}
    \caption{Summary of outlier fractions, NMADs and bias achieved for all color-shifting and augmentation cases. We have abbreviated redshift augmentation as `$z$', magnitude augmentation as `Mag' and color augmentation as `col'. Values in the parentheses in the outlier fraction columns are the catastrophic outliers. The bolded values correspond to the best performing augmentation case.}
    \label{tab:summary}
\end{table*}

There are three double feature combinations possible: magnitude+redshift, color+redshift and color+magnitude. Since the training sample creates a compact shape in color-magnitude space, we fit two lines to mimic the shape of the top of the color-magnitude distribution of the training sample. We then choose objects above and to the left of this region, see the dashed line and solid arrows in the right panel of Figure \ref{fig:nonrep_training}. 

The training sample does not have a simple shape in either color-redshift or magnitide-redshift space. For these feature combinations, we use the intersection of the selection requirements for each feature rather than the union; for example, the magnitude+redshift augmentation selects Buzzard galaxies with both $i$-mag $> 23$ and $z_{true} > 1.0$. The intersection worked mildly better than the union. 

When using multiple features to select Buzzard galaxies for augmentation, the magnitude shifted Buzzard sample produced the best results. Results for the unshifted and flowed Buzzard catalogs are also listed in Table \ref{tab:summary}.

Similar to the single feature augmentation, the double feature combinations that include redshift as a selection criterion perform better than combinations without redshift. Magnitude+Redshift and Color+Redshift combinations perform virtually identically for the magnitude shifted Buzzard catalog, with outlier fractions of 0.259 and 0.258, respectivly. This corresponds to a ratio to the unaugmented results of 0.54, and a 65\% recovery of the fully representative case. Both produce an NMAD of 0.086, which corresponds to a ratio to the unaugmented results of 0.452 and a 78\% recovery towards the fully representative case. The color+redshift augmented training sample, using the magnitude shifted Buzzard catalog, and the resulting photo-$z$ estimates are shown in the middle panel of Figure \ref{fig:best_results}.

When compared to the outlier fraction for the redshift augmentation using the magnitude shifted catalog (see Table \ref{tab:summary}), adding color or magnitude provides a small improvement. The color+magnitude combination also provides a small improvement over either color or magnitude alone in most cases.

\subsection{Color+Magnitude+Redshift Augmentation} \label{best_results}
Finally, we use all three features to select Buzzard galaxies for augmentation. We use the intersection of the Color+Magnitude and redshift selection criteria as this provides better results than the union. This is likely because we always augment the training sample with 10,000 galaxies; in this case, the intersection most efficiently probes the feature space not covered by the DC2 training sample. The color+magnitude+redshift augmented training sample, using the shifted magnitude Buzzard catalog, is shown in the top right of Figure \ref{fig:best_results}, and the post augmentation redshift distribution is shown as the black dot-dashed line in Figure \ref{fig:nonrep_redshifts}.

As with the double feature augmentation cases, the simple magnitude shifted Buzzard catalog produced the best results, and we show that case in the right panel of Figure \ref{fig:best_results}.

This augmentation case produces the lowest outlier fraction and NMAD out of all combinations of color shifting and augmentation features. At an outlier fraction of 0.245, which is a ratio of 0.51 to the unaugmented case, and recovers $69\%$ of the degradation resulting from the non-representative training sample. We achieve an NMAD of 0.084, a ratio of 0.44 to the unaugmented case,  and an $80\%$ recovery of the degradation in NMAD compared to the fully representative training sample. We also achieve 8 times less bias with this augmentation than in the unaugmented case (see Table \ref{tab:summary}).

It can be seen that the augmented training samples using the shifted magnitudes Buzzard catalog have \textit{i}-band magnitudes that extend fainter than the application sample. We tried imposing an $i<26$ cut on the magnitude shifted Buzzard catalog before selecting galaxies for augmenting, but this produces slightly worse photo-$z$ estimates. We suspect this is because FlexZBoost is a conditional density estimator, and a hard cut-off in the magnitudes makes it difficult for FlexZBoost to estimate the density at the cut-off magnitude. Allowing the magnitudes to drop off naturally makes it easier for FlexZBoost to estimate the density at $i=26$, resulting in better photo-$z$ estimates even though the magnitude range extends farther than required to match the application sample.

\subsection{Comparison to TPZ}
\textbf{To test whether the results of augmentation depend strongly on the machine learning method used, we also estimated photo-$z$'s using the RAIL implementation of the code Trees for Photo-Z (TPZ, \citealt{tpz}), which uses a random forest method. TPZ produces similar results when no training sample augmentation is performed; the photo-$z$ statistics of (outlier fraction, catastrophic outlier fraction, NMAD and bias) for the unaugmented TPZ case is (0.51, 0.20, 0.22, -0.13) compared to (0.48, 0.21, 0.19, -0.12) for the unaugmented FlexZBoost case. The best performing augmentation case, using the magnitude shifted Buzzard sample and augmenting with color+magnitude+redshift, produced photo-$z$ statistics of (0.32, 0.04, 0.11, -0.014) for TPZ, a smaller but still highly significant improvement over the unaugmented case when compared to the FlexZBoost results (see Table \ref{tab:summary}).}  

\section{Conclusions}
Large imaging surveys, such as LSST, will not have access to representative training samples for estimating photometric redshifts. When estimating photo-$z$'s using a realistically non-representative training sample, the outlier fraction reaches nearly $50\%$, almost a factor of 3.5 worse than when using a representative training sample, and a similar increase in scatter as probed by the NMAD. Obtaining new spectroscopic samples of dim galaxies cannot solve this problem alone, as it is not feasible to obtain a large enough sample by the time LSST is expected to see first light. Training sample augmentation is an easy way to improve photo-$z$ estimates without requiring additional spectroscopic samples of dim galaxies. 

We used the DESC DC2 simulation as a stand-in for eventual LSST data, and investigated how augmenting a realistically non-representative training sample with simulated galaxies from Buzzard can improve the photo-$z$ estimates. Even a relatively simple augmentation process of selecting simulated galaxies with redshifts higher than those present in the training sample can recover $65\%$ of the degradation in the outlier fraction when compared to a fully representative training sample.

Shifting the photometry of the galaxy catalog used for augmenting the training sample can improve the results further. We shifted all Buzzard magnitudes so the median magnitude in each band matches the median magnitudes of the DC2 application sample. We then select galaxies for augmentation in regions of color-magnitude-redshift space not covered by the training sample. The resulting photo-$z$ estimates, shown in the right panel of 
Figure~\ref{fig:best_results} and Table~\ref{tab:summary}, have an outlier fraction below $25\%$, NMAD of 0.084, and bias of -0.014, representing a nearly $50\%$ reduction over the unaugmented photo-$z$ outlier fraction, $56\%$ reduction in NMAD, and factor of 8 reduction in bias. Augmentation has recovered $69\%$ of the degradation in outlier fraction compared to the fully representative case, $80\%$ of the degradation in NMAD, and $88\%$ of the degradation in bias. With these results in mind, it is clear that training sample augmentation should be considered in the photo-$z$ pipeline for large galaxy surveys, including LSST.

Since redshift seems to be the most important feature for augmentation, the Buzzard catalog is not the best-case scenario for augmentation given how few galaxies there are at $z > 1.0$. There are almost no Buzzard objects at $z > 1.5$, while there are still many application sample objects in that redshift range. When using augmentation for real data, choosing a simulation with sufficient redshift coverage, such as the DC2 simulation should produce better results than were achieved here.

While we leave extensions of this method to real data to future work, we discuss below a few possible avenues to explore training sample augmentation for real data. One could, for example, use the deep photo-z catalogs from COSMOS2020 \citep{cosmos2020} as a truth catalog, where the COSMOS2020 photo-z catalog would take the place of the DC2 catalog used in this work. It could also be worthwile to explore using a deep photo-z catalog such as COSMO2020 as the augmentation catalog, in place of the Buzzard catalog.

Self-organizing maps (SOMs) have also been used to correct for spectroscopic incompleteness of training samples in DES 
(see, for example, \citealt{des_y3_photoz, pheno-z_des}) and KiDS \citep{kids_som}. Assigning training and application samples to SOM cells can be useful for identifying under-represented regions of photometry in the training sample that can be targeted for spectroscopic followup \citep{som_followup}. Future efforts could use the SOM methodology to identify photometry regions to target with augmented training samples.

We have assumed in this work that the outlier fraction and NMAD are good indicators of photo-$z$ quality. While this is a reasonable assumption, a full quantification of the improvements provided by augmentation would come from using the photo-$z$ estimates to do a full cosmological parameter estimation. For augmentation to truly be useful, it should result in better cosmological parameter estimates than the unaugmented photo-$z$'s. This analysis will be presented in a forthcoming paper.

\section{Acknowledgements}
This paper has undergone internal review by the LSST Dark Energy Science Collaboration. The internal reviewers were Huan Lin and Eve Kovacs. The authors thank the internal reviewers for their valuable comments. 

IM and EG acknowledge support for this research from the LSST Corporation via grant \#2021-42. IM and EG also acknowledge support from the U.S. Department of Energy, Office of Science, Office of High Energy Physics Cosmic Frontier Research program under Award Number DE-SC0010008. JFC acknowledges support from the U.S. Department of Energy, Office of Science, Office of High Energy Physics Cosmic Frontier Research program under Award Number DE-SC0011665.  BHA acknowledges support by the National Science Foundation under
Award Number AST-2009251. AIM acknowledges the support of Schmidt Sciences.

Author contributions are as follows. IM performed the analysis and wrote the majority of the paper. EG advised IM, suggested approaches, and provided feedback on the text. JFC suggested and advised on methodology for the normalizing flow, and provided feedback on manuscript. BHA provided guidance on bias discussions and feedback on the manuscript. AIM designed and led the development of the RAIL software. SS advised on the use of FlexZBoost and feedback on the manuscript. 

The DESC acknowledges ongoing support from the Institut National de 
Physique Nucl\'eaire et de Physique des Particules in France; the 
Science \& Technology Facilities Council in the United Kingdom; and the
Department of Energy, the National Science Foundation, and the LSST 
Corporation in the United States.  DESC uses resources of the IN2P3 
Computing Center (CC-IN2P3--Lyon/Villeurbanne - France) funded by the 
Centre National de la Recherche Scientifique; the National Energy 
Research Scientific Computing Center, a DOE Office of Science User 
Facility supported by the Office of Science of the U.S.\ Department of
Energy under Contract No.\ DE-AC02-05CH11231; STFC DiRAC HPC Facilities, 
funded by UK BEIS National E-infrastructure capital grants; and the UK 
particle physics grid, supported by the GridPP Collaboration.  This 
work was performed in part under DOE Contract DE-AC02-76SF00515.

\bibliographystyle{aasjournal}
\bibliography{biblio}

\begin{thebibliography}{}
\expandafter\ifx\csname natexlab\endcsname\relax\def\natexlab#1{#1}\fi
\providecommand{\url}[1]{\href{#1}{#1}}
\providecommand{\dodoi}[1]{doi:~\href{http://doi.org/#1}{\nolinkurl{#1}}}
\providecommand{\doeprint}[1]{\href{http://ascl.net/#1}{\nolinkurl{http://ascl.net/#1}}}
\providecommand{\doarXiv}[1]{\href{https://arxiv.org/abs/#1}{\nolinkurl{https://arxiv.org/abs/#1}}}

\bibitem[{Abbott {et~al.}(2018)Abbott, Abdalla, Alarcon, Aleksi\ifmmode~\acute{c}\else \'{c}\fi{}, Allam, Allen, Amara, Annis, Asorey, Avila, Bacon, Balbinot, Banerji, Banik, Barkhouse, Baumer, Baxter, Bechtol, Becker, Benoit-L\'evy, Benson, Bernstein, Bertin, Blazek, Bridle, Brooks, Brout, Buckley-Geer, Burke, Busha, Campos, Capozzi, Carnero~Rosell, Carrasco~Kind, Carretero, Castander, Cawthon, Chang, Chen, Childress, Choi, Conselice, Crittenden, Crocce, Cunha, D'Andrea, da~Costa, Das, Davis, Davis, De~Vicente, DePoy, DeRose, Desai, Diehl, Dietrich, Dodelson, Doel, Drlica-Wagner, Eifler, Elliott, Elsner, Elvin-Poole, Estrada, Evrard, Fang, Fernandez, Fert\'e, Finley, Flaugher, Fosalba, Friedrich, Frieman, Garc\'{\i}a-Bellido, Garcia-Fernandez, Gatti, Gaztanaga, Gerdes, Giannantonio, Gill, Glazebrook, Goldstein, Gruen, Gruendl, Gschwend, Gutierrez, Hamilton, Hartley, Hinton, Honscheid, Hoyle, Huterer, Jain, James, Jarvis, Jeltema, Johnson, Johnson, Kacprzak, Kent, Kim, King, Kirk, Kokron, Kovacs, Krause,
  Krawiec, Kremin, Kuehn, Kuhlmann, Kuropatkin, Lacasa, Lahav, Li, Liddle, Lidman, Lima, Lin, MacCrann, Maia, Makler, Manera, March, Marshall, Martini, McMahon, Melchior, Menanteau, Miquel, Miranda, Mudd, Muir, M\"oller, Neilsen, Nichol, Nord, Nugent, Ogando, Palmese, Peacock, Peiris, Peoples, Percival, Petravick, Plazas, Porredon, Prat, Pujol, Rau, Refregier, Ricker, Roe, Rollins, Romer, Roodman, Rosenfeld, Ross, Rozo, Rykoff, Sako, Salvador, Samuroff, S\'anchez, Sanchez, Santiago, Scarpine, Schindler, Scolnic, Secco, Serrano, Sevilla-Noarbe, Sheldon, Smith, Smith, Smith, Soares-Santos, Sobreira, Suchyta, Tarle, Thomas, Troxel, Tucker, Tucker, Uddin, Varga, Vielzeuf, Vikram, Vivas, Walker, Wang, Wechsler, Weller, Wester, Wolf, Yanny, Yuan, Zenteno, Zhang, Zhang, \& Zuntz}]{DES}
Abbott, T. M.~C., Abdalla, F.~B., Alarcon, A., {et~al.} 2018, Phys. Rev. D, 98, 043526, \dodoi{10.1103/PhysRevD.98.043526}

\bibitem[{Abolfathi {et~al.}(2021)Abolfathi, Alonso, Armstrong, Aubourg, Awan, Babuji, Bauer, Bean, Beckett, Biswas, Bogart, Boutigny, Chard, Chiang, Claver, Cohen-Tanugi, Combet, Connolly, Daniel, Digel, Drlica-Wagner, Dubois, Gangler, Gawiser, Glanzman, Gris, Habib, Hearin, Heitmann, Hernandez, Hlo{\v{z}}ek, Hollowed, Ishak, Ivezi{\'{c}}, Jarvis, Jha, Kahn, Kalmbach, Kelly, Kovacs, Korytov, Krughoff, Lage, Lanusse, Larsen, Guillou, Li, Longley, Lupton, Mandelbaum, Mao, Marshall, Meyers, Moniez, Morrison, Nomerotski, O'Connor, Park, Park, Peloton, Perrefort, Perry, Plaszczynski, Pope, Rasmussen, Reil, Roodman, Rykoff, S{\'{a}}nchez, Schmidt, Scolnic, Stubbs, Tyson, Uram, Villarreal, Walter, Wiesner, Wood-Vasey, \& Zuntz}]{DC2}
Abolfathi, B., Alonso, D., Armstrong, R., {et~al.} 2021, The Astrophysical Journal Supplement Series, 253, 31, \dodoi{10.3847/1538-4365/abd62c}

\bibitem[{Aihara {et~al.}(2017)Aihara, Arimoto, Armstrong, Arnouts, Bahcall, Bickerton, Bosch, Bundy, Capak, Chan, Chiba, Coupon, Egami, Enoki, Finet, Fujimori, Fujimoto, Furusawa, Furusawa, Goto, Goulding, Greco, Greene, Gunn, Hamana, Harikane, Hashimoto, Hattori, Hayashi, Hayashi, He{\l}miniak, Higuchi, Hikage, Ho, Hsieh, Huang, Huang, Ikeda, Imanishi, Inoue, Iwasawa, Iwata, Jaelani, Jian, Kamata, Karoji, Kashikawa, Katayama, Kawanomoto, Kayo, Koda, Koike, Kojima, Komiyama, Konno, Koshida, Koyama, Kusakabe, Leauthaud, Lee, Lin, Lin, Lupton, Mandelbaum, Matsuoka, Medezinski, Mineo, Miyama, Miyatake, Miyazaki, Momose, More, More, Moritani, Moriya, Morokuma, Mukae, Murata, Murayama, Nagao, Nakata, Niida, Niikura, Nishizawa, Obuchi, Oguri, Oishi, Okabe, Okamoto, Okura, Ono, Onodera, Onoue, Osato, Ouchi, Price, Pyo, Sako, Sawicki, Shibuya, Shimasaku, Shimono, Shirasaki, Silverman, Simet, Speagle, Spergel, Strauss, Sugahara, Sugiyama, Suto, Suyu, Suzuki, Tait, Takada, Takata, Tamura, Tanaka, Tanaka, Tanaka,
  Tanaka, Terai, Terashima, Toba, Tominaga, Toshikawa, Turner, Uchida, Uchiyama, Umetsu, Uraguchi, Urata, Usuda, Utsumi, Wang, Wang, Wong, Yabe, Yamada, Yamanoi, Yasuda, Yeh, Yonehara, \& Yuma}]{hsc_overview}
Aihara, H., Arimoto, N., Armstrong, R., {et~al.} 2017, Publications of the Astronomical Society of Japan, 70, S4, \dodoi{10.1093/pasj/psx066}

\bibitem[{{Aihara} {et~al.}(2019){Aihara}, {AlSayyad}, {Ando}, {Armstrong}, {Bosch}, {Egami}, {Furusawa}, {Furusawa}, {Goulding}, {Harikane}, {Hikage}, {Ho}, {Hsieh}, {Huang}, {Ikeda}, {Imanishi}, {Ito}, {Iwata}, {Jaelani}, {Kakuma}, {Kawana}, {Kikuta}, {Kobayashi}, {Koike}, {Komiyama}, {Li}, {Liang}, {Lin}, {Luo}, {Lupton}, {Lust}, {MacArthur}, {Matsuoka}, {Mineo}, {Miyatake}, {Miyazaki}, {More}, {Murata}, {Namiki}, {Nishizawa}, {Oguri}, {Okabe}, {Okamoto}, {Okura}, {Ono}, {Onodera}, {Onoue}, {Osato}, {Ouchi}, {Shibuya}, {Strauss}, {Sugiyama}, {Suto}, {Takada}, {Takagi}, {Takata}, {Takita}, {Tanaka}, {Terai}, {Toba}, {Uchiyama}, {Utsumi}, {Wang}, {Wang}, \& {Yamada}}]{hsc_dr2}
{Aihara}, H., {AlSayyad}, Y., {Ando}, M., {et~al.} 2019, \pasj, 71, 114, \dodoi{10.1093/pasj/psz103}

\bibitem[{Akeson {et~al.}(2019)Akeson, Armus, Bachelet, Bailey, Bartusek, Bellini, Benford, Bennett, Bhattacharya, Bohlin, Boyer, Bozza, Bryden, Novati, Carpenter, Casertano, Choi, Content, Dayal, Dressler, Doré, Fall, Fan, Fang, Filippenko, Finkelstein, Foley, Furlanetto, Kalirai, Gaudi, Gilbert, Girard, Grady, Greene, Guhathakurta, Heinrich, Hemmati, Hendel, Henderson, Henning, Hirata, Ho, Huff, Hutter, Jansen, Jha, Johnson, Jones, Kasdin, Kelly, Kirshner, Koekemoer, Kruk, Lewis, Macintosh, Madau, Malhotra, Mandel, Massara, Masters, McEnery, McQuinn, Melchior, Melton, Mennesson, Peeples, Penny, Perlmutter, Pisani, Plazas, Poleski, Postman, Ranc, Rauscher, Rest, Roberge, Robertson, Rodney, Rhoads, Rhodes, Ryan, Sahu, Sand, Scolnic, Seth, Shvartzvald, Siellez, Smith, Spergel, Stassun, Street, Strolger, Szalay, Trauger, Troxel, Turnbull, van~der Marel, von~der Linden, Wang, Weinberg, Williams, Windhorst, Wollack, Wu, Yee, \& Zimmerman}]{roman}
Akeson, R., Armus, L., Bachelet, E., {et~al.} 2019, The Wide Field Infrared Survey Telescope: 100 Hubbles for the 2020s,  arXiv, \dodoi{10.48550/ARXIV.1902.05569}

\bibitem[{{Beck} {et~al.}(2017){Beck}, {Lin}, {Ishida}, {Gieseke}, {de Souza}, {Costa-Duarte}, {Hattab}, \& {Krone-Martins}}]{beck2017}
{Beck}, R., {Lin}, C.~A., {Ishida}, E.~E.~O., {et~al.} 2017, \mnras, 468, 4323, \dodoi{10.1093/mnras/stx687}

\bibitem[{Behroozi {et~al.}(2019)Behroozi, Wechsler, Hearin, \& Conroy}]{universemachine}
Behroozi, P., Wechsler, R.~H., Hearin, A.~P., \& Conroy, C. 2019, Monthly Notices of the Royal Astronomical Society, 488, 3143–3194, \dodoi{10.1093/mnras/stz1182}

\bibitem[{Benson(2012)}]{galacticus}
Benson, A.~J. 2012, New Astronomy, 17, 175–197, \dodoi{10.1016/j.newast.2011.07.004}

\bibitem[{Bird {et~al.}(2021)Bird, Pritchard, Fratini, Ekárt, \& Faria}]{synthetic_augmentation}
Bird, J., Pritchard, M., Fratini, A., Ekárt, A., \& Faria, D. 2021, IEEE Robotics and Automation Letters, 6, 3498, \dodoi{10.1109/LRA.2021.3056355}

\bibitem[{Bloice {et~al.}(2017)Bloice, Stocker, \& Holzinger}]{image_augmentation}
Bloice, M.~D., Stocker, C., \& Holzinger, A. 2017, Journal of Open Source Software, 2, 432, \dodoi{10.21105/joss.00432}

\bibitem[{Broussard \& Gawiser(2021)}]{Broussard_2021}
Broussard, A., \& Gawiser, E. 2021, The Astrophysical Journal, 922, 153, \dodoi{10.3847/1538-4357/ac2147}

\bibitem[{{Buchs} {et~al.}(2019){Buchs}, {Davis}, {Gruen}, {DeRose}, {Alarcon}, {Bernstein}, {S{\'a}nchez}, {Myles}, {Roodman}, {Allen}, {Amon}, {Choi}, {Masters}, {Miquel}, {Troxel}, {Wechsler}, {Abbott}, {Annis}, {Avila}, {Bechtol}, {Bridle}, {Brooks}, {Buckley-Geer}, {Burke}, {Carnero Rosell}, {Carrasco Kind}, {Carretero}, {Castander}, {Cawthon}, {D'Andrea}, {da Costa}, {De Vicente}, {Desai}, {Diehl}, {Doel}, {Drlica-Wagner}, {Eifler}, {Evrard}, {Flaugher}, {Fosalba}, {Frieman}, {Garc{\'\i}a-Bellido}, {Gaztanaga}, {Gruendl}, {Gschwend}, {Gutierrez}, {Hartley}, {Hollowood}, {Honscheid}, {James}, {Kuehn}, {Kuropatkin}, {Lima}, {Lin}, {Maia}, {March}, {Marshall}, {Melchior}, {Menanteau}, {Ogando}, {Plazas}, {Rykoff}, {Sanchez}, {Scarpine}, {Serrano}, {Sevilla-Noarbe}, {Smith}, {Soares-Santos}, {Sobreira}, {Suchyta}, {Swanson}, {Tarle}, {Thomas}, {Vikram}, \& {DES Collaboration}}]{pheno-z_des}
{Buchs}, R., {Davis}, C., {Gruen}, D., {et~al.} 2019, \mnras, 489, 820, \dodoi{10.1093/mnras/stz2162}

\bibitem[{Carrasco~Kind \& Brunner(2013)}]{tpz}
Carrasco~Kind, M., \& Brunner, R.~J. 2013, Monthly Notices of the Royal Astronomical Society, 432, 1483, \dodoi{10.1093/mnras/stt574}

\bibitem[{{Conroy} {et~al.}(2009){Conroy}, {Gunn}, \& {White}}]{fsps}
{Conroy}, C., {Gunn}, J.~E., \& {White}, M. 2009, \apj, 699, 486, \dodoi{10.1088/0004-637X/699/1/486}

\bibitem[{Crenshaw {et~al.}(2023)Crenshaw, Yan, \& Doster}]{pzflow}
Crenshaw, J.~F., Yan, Z., \& Doster, V. 2023, jfcrenshaw/pzflow: v3.1.1, v3.1.1,  Zenodo, \dodoi{https://doi.org/10.5281/zenodo.7843901}

\bibitem[{{Dalmasso} {et~al.}(2020){Dalmasso}, {Pospisil}, {Lee}, {Izbicki}, {Freeman}, \& {Malz}}]{fzboost2020}
{Dalmasso}, N., {Pospisil}, T., {Lee}, A.~B., {et~al.} 2020, Astronomy and Computing, 30, 100362, \dodoi{10.1016/j.ascom.2019.100362}

\bibitem[{{de Jong} {et~al.}(2019){de Jong}, {Agertz}, {Berbel}, {Aird}, {Alexander}, {Amarsi}, {Anders}, {Andrae}, {Ansarinejad}, {Ansorge}, {Antilogus}, {Anwand-Heerwart}, {Arentsen}, {Arnadottir}, {Asplund}, {Auger}, {Azais}, {Baade}, {Baker}, {Baker}, {Balbinot}, {Baldry}, {Banerji}, {Barden}, {Barklem}, {Barth{\'e}l{\'e}my-Mazot}, {Battistini}, {Bauer}, {Bell}, {Bellido-Tirado}, {Bellstedt}, {Belokurov}, {Bensby}, {Bergemann}, {Bestenlehner}, {Bielby}, {Bilicki}, {Blake}, {Bland-Hawthorn}, {Boeche}, {Boland}, {Boller}, {Bongard}, {Bongiorno}, {Bonifacio}, {Boudon}, {Brooks}, {Brown}, {Brown}, {Br{\"u}ggen}, {Brynnel}, {Brzeski}, {Buchert}, {Buschkamp}, {Caffau}, {Caillier}, {Carrick}, {Casagrande}, {Case}, {Casey}, {Cesarini}, {Cescutti}, {Chapuis}, {Chiappini}, {Childress}, {Christlieb}, {Church}, {Cioni}, {Cluver}, {Colless}, {Collett}, {Comparat}, {Cooper}, {Couch}, {Courbin}, {Croom}, {Croton}, {Daguis{\'e}}, {Dalton}, {Davies}, {Davis}, {de Laverny}, {Deason}, {Dionies}, {Disseau}, {Doel},
  {D{\"o}scher}, {Driver}, {Dwelly}, {Eckert}, {Edge}, {Edvardsson}, {Youssoufi}, {Elhaddad}, {Enke}, {Erfanianfar}, {Farrell}, {Fechner}, {Feiz}, {Feltzing}, {Ferreras}, {Feuerstein}, {Feuillet}, {Finoguenov}, {Ford}, {Fotopoulou}, {Fouesneau}, {Frenk}, {Frey}, {Gaessler}, {Geier}, {Gentile Fusillo}, {Gerhard}, {Giannantonio}, {Giannone}, {Gibson}, {Gillingham}, {Gonz{\'a}lez-Fern{\'a}ndez}, {Gonzalez-Solares}, {Gottloeber}, {Gould}, {Grebel}, {Gueguen}, {Guiglion}, {Haehnelt}, {Hahn}, {Hansen}, {Hartman}, {Hauptner}, {Hawkins}, {Haynes}, {Haynes}, {Heiter}, {Helmi}, {Aguayo}, {Hewett}, {Hinton}, {Hobbs}, {Hoenig}, {Hofman}, {Hook}, {Hopgood}, {Hopkins}, {Hourihane}, {Howes}, {Howlett}, {Huet}, {Irwin}, {Iwert}, {Jablonka}, {Jahn}, {Jahnke}, {Jarno}, {Jin}, {Jofre}, {Johl}, {Jones}, {J{\"o}nsson}, {Jordan}, {Karovicova}, {Khalatyan}, {Kelz}, {Kennicutt}, {King}, {Kitaura}, {Klar}, {Klauser}, {Kneib}, {Koch}, {Koposov}, {Kordopatis}, {Korn}, {Kosmalski}, {Kotak}, {Kovalev}, {Kreckel}, {Kripak}, {Krumpe},
  {Kuijken}, {Kunder}, {Kushniruk}, {Lam}, {Lamer}, {Laurent}, {Lawrence}, {Lehmitz}, {Lemasle}, {Lewis}, {Li}, {Lidman}, {Lind}, {Liske}, {Lizon}, {Loveday}, {Ludwig}, {McDermid}, {Maguire}, {Mainieri}, {Mali}, {Mandel}, {Mandel}, {Mannering}, {Martell}, {Martinez Delgado}, {Matijevic}, {McGregor}, {McMahon}, {McMillan}, {Mena}, {Merloni}, {Meyer}, {Michel}, {Micheva}, {Migniau}, {Minchev}, {Monari}, {Muller}, {Murphy}, {Muthukrishna}, {Nandra}, {Navarro}, {Ness}, {Nichani}, {Nichol}, {Nicklas}, {Niederhofer}, {Norberg}, {Obreschkow}, {Oliver}, {Owers}, {Pai}, {Pankratow}, {Parkinson}, {Paschke}, {Paterson}, {Pecontal}, {Parry}, {Phillips}, {Pillepich}, {Pinard}, {Pirard}, {Piskunov}, {Plank}, {Pl{\"u}schke}, {Pons}, {Popesso}, {Power}, {Pragt}, {Pramskiy}, {Pryer}, {Quattri}, {Queiroz}, {Quirrenbach}, {Rahurkar}, {Raichoor}, {Ramstedt}, {Rau}, {Recio-Blanco}, {Reiss}, {Renaud}, {Revaz}, {Rhode}, {Richard}, {Richter}, {Rix}, {Robotham}, {Roelfsema}, {Romaniello}, {Rosario}, {Rothmaier}, {Roukema}, {Ruchti},
  {Rupprecht}, {Rybizki}, {Ryde}, {Saar}, {Sadler}, {Sahl{\'e}n}, {Salvato}, {Sassolas}, {Saunders}, {Saviauk}, {Sbordone}, {Schmidt}, {Schnurr}, {Scholz}, {Schwope}, {Seifert}, {Shanks}, {Sheinis}, {Sivov}, {Sk{\'u}lad{\'o}ttir}, {Smartt}, {Smedley}, {Smith}, {Smith}, {Sorce}, {Spitler}, {Starkenburg}, {Steinmetz}, {Stilz}, {Storm}, {Sullivan}, {Sutherland}, {Swann}, {Tamone}, {Taylor}, {Teillon}, {Tempel}, {ter Horst}, {Thi}, {Tolstoy}, {Trager}, {Traven}, {Tremblay}, {Tresse}, {Valentini}, {van de Weygaert}, {van den Ancker}, {Veljanoski}, {Venkatesan}, {Wagner}, {Wagner}, {Walcher}, {Waller}, {Walton}, {Wang}, {Winkler}, {Wisotzki}, {Worley}, {Worseck}, {Xiang}, {Xu}, {Yong}, {Zhao}, {Zheng}, {Zscheyge}, \& {Zucker}}]{4most}
{de Jong}, R.~S., {Agertz}, O., {Berbel}, A.~A., {et~al.} 2019, The Messenger, 175, 3, \dodoi{10.18727/0722-6691/5117}

\bibitem[{DeRose {et~al.}(2019)DeRose, Wechsler, Becker, Busha, Rykoff, MacCrann, Erickson, Evrard, Kravtsov, Gruen, Allam, Avila, Bridle, Brooks, Buckley-Geer, Rosell, Kind, Carretero, Castander, Cawthon, Crocce, da~Costa, Davis, Vicente, Dietrich, Doel, Drlica-Wagner, Fosalba, Frieman, Garcia-Bellido, Gutierrez, Hartley, Hollowood, Hoyle, James, Krause, Kuehn, Kuropatkin, Lima, Maia, Menanteau, Miller, Miquel, Ogando, Malagón, Romer, Sanchez, Schindler, Serrano, Sevilla-Noarbe, Smith, Suchyta, Swanson, Tarle, \& Vikram}]{buzzard}
DeRose, J., Wechsler, R.~H., Becker, M.~R., {et~al.} 2019, The Buzzard Flock: Dark Energy Survey Synthetic Sky Catalogs.
\newblock \doarXiv{1901.02401}

\bibitem[{{Euclid Collaboration} {et~al.}(2022){Euclid Collaboration}, {Scaramella}, {Amiaux}, {Mellier}, {Burigana}, {Carvalho}, {Cuillandre}, {Da Silva}, {Derosa}, {Dinis}, {Maiorano}, {Maris}, {Tereno}, {Laureijs}, {Boenke}, {Buenadicha}, {Dupac}, {Gaspar Venancio}, {G{\'o}mez-{\'A}lvarez}, {Hoar}, {Lorenzo Alvarez}, {Racca}, {Saavedra-Criado}, {Schwartz}, {Vavrek}, {Schirmer}, {Aussel}, {Azzollini}, {Cardone}, {Cropper}, {Ealet}, {Garilli}, {Gillard}, {Granett}, {Guzzo}, {Hoekstra}, {Jahnke}, {Kitching}, {Maciaszek}, {Meneghetti}, {Miller}, {Nakajima}, {Niemi}, {Pasian}, {Percival}, {Pottinger}, {Sauvage}, {Scodeggio}, {Wachter}, {Zacchei}, {Aghanim}, {Amara}, {Auphan}, {Auricchio}, {Awan}, {Balestra}, {Bender}, {Bodendorf}, {Bonino}, {Branchini}, {Brau-Nogue}, {Brescia}, {Candini}, {Capobianco}, {Carbone}, {Carlberg}, {Carretero}, {Casas}, {Castander}, {Castellano}, {Cavuoti}, {Cimatti}, {Cledassou}, {Congedo}, {Conselice}, {Conversi}, {Copin}, {Corcione}, {Costille}, {Courbin}, {Degaudenzi}, {Douspis},
  {Dubath}, {Duncan}, {Dusini}, {Farrens}, {Ferriol}, {Fosalba}, {Fourmanoit}, {Frailis}, {Franceschi}, {Franzetti}, {Fumana}, {Gillis}, {Giocoli}, {Grazian}, {Grupp}, {Haugan}, {Holmes}, {Hormuth}, {Hudelot}, {Kermiche}, {Kiessling}, {Kilbinger}, {Kohley}, {Kubik}, {K{\"u}mmel}, {Kunz}, {Kurki-Suonio}, {Lahav}, {Ligori}, {Lilje}, {Lloro}, {Mansutti}, {Marggraf}, {Markovic}, {Marulli}, {Massey}, {Maurogordato}, {Melchior}, {Merlin}, {Meylan}, {Mohr}, {Moresco}, {Morin}, {Moscardini}, {Munari}, {Nichol}, {Padilla}, {Paltani}, {Peacock}, {Pedersen}, {Pettorino}, {Pires}, {Poncet}, {Popa}, {Pozzetti}, {Raison}, {Rebolo}, {Rhodes}, {Rix}, {Roncarelli}, {Rossetti}, {Saglia}, {Schneider}, {Schrabback}, {Secroun}, {Seidel}, {Serrano}, {Sirignano}, {Sirri}, {Skottfelt}, {Stanco}, {Starck}, {Tallada-Cresp{\'\i}}, {Tavagnacco}, {Taylor}, {Teplitz}, {Toledo-Moreo}, {Torradeflot}, {Trifoglio}, {Valentijn}, {Valenziano}, {Verdoes Kleijn}, {Wang}, {Welikala}, {Weller}, {Wetzstein}, {Zamorani}, {Zoubian}, {Andreon},
  {Baldi}, {Bardelli}, {Boucaud}, {Camera}, {Di Ferdinando}, {Fabbian}, {Farinelli}, {Galeotta}, {Graci{\'a}-Carpio}, {Maino}, {Medinaceli}, {Mei}, {Neissner}, {Polenta}, {Renzi}, {Romelli}, {Rosset}, {Sureau}, {Tenti}, {Vassallo}, {Zucca}, {Baccigalupi}, {Balaguera-Antol{\'\i}nez}, {Battaglia}, {Biviano}, {Borgani}, {Bozzo}, {Cabanac}, {Cappi}, {Casas}, {Castignani}, {Colodro-Conde}, {Coupon}, {Courtois}, {Cuby}, {de la Torre}, {Desai}, {Dole}, {Fabricius}, {Farina}, {Ferreira}, {Finelli}, {Flose-Reimberg}, {Fotopoulou}, {Ganga}, {Gozaliasl}, {Hook}, {Keihanen}, {Kirkpatrick}, {Liebing}, {Lindholm}, {Mainetti}, {Martinelli}, {Martinet}, {Maturi}, {McCracken}, {Metcalf}, {Morgante}, {Nightingale}, {Nucita}, {Patrizii}, {Potter}, {Riccio}, {S{\'a}nchez}, {Sapone}, {Schewtschenko}, {Schultheis}, {Scottez}, {Teyssier}, {Tutusaus}, {Valiviita}, {Viel}, {Vriend}, \& {Whittaker}}]{euclid}
{Euclid Collaboration}, {Scaramella}, R., {Amiaux}, J., {et~al.} 2022, \aap, 662, A112, \dodoi{10.1051/0004-6361/202141938}

\bibitem[{{Flaugher} \& {Bebek}(2014)}]{desi}
{Flaugher}, B., \& {Bebek}, C. 2014, in Society of Photo-Optical Instrumentation Engineers (SPIE) Conference Series, Vol. 9147, Ground-based and Airborne Instrumentation for Astronomy V, ed. S.~K. {Ramsay}, I.~S. {McLean}, \& H.~{Takami}, 91470S, \dodoi{10.1117/12.2057105}

\bibitem[{Hearin {et~al.}(2020)Hearin, Korytov, Kovacs, Benson, Aung, Bradshaw, \& Campbell}]{galsampler}
Hearin, A., Korytov, D., Kovacs, E., {et~al.} 2020, Monthly Notices of the Royal Astronomical Society, 495, 5040–5051, \dodoi{10.1093/mnras/staa1495}

\bibitem[{{Heitmann} {et~al.}(2019){Heitmann}, {Finkel}, {Pope}, {Morozov}, {Frontiere}, {Habib}, {Rangel}, {Uram}, {Korytov}, {Child}, {Flender}, {Insley}, \& {Rizzi}}]{outer_rim}
{Heitmann}, K., {Finkel}, H., {Pope}, A., {et~al.} 2019, \apjs, 245, 16, \dodoi{10.3847/1538-4365/ab4da1}

\bibitem[{Heymans {et~al.}(2021)Heymans, Tröster, Asgari, Blake, Hildebrandt, Joachimi, Kuijken, Lin, S{\'{a} }nchez, van~den Busch, Wright, Amon, Bilicki, de~Jong, Crocce, Dvornik, Erben, Fortuna, Getman, Giblin, Glazebrook, Hoekstra, Joudaki, Kannawadi, Köhlinger, Lidman, Miller, Napolitano, Parkinson, Schneider, Shan, Valentijn, Kleijn, \& Wolf}]{kids}
Heymans, C., Tröster, T., Asgari, M., {et~al.} 2021, Astronomy {\&} Astrophysics, 646, A140, \dodoi{10.1051/0004-6361/202039063}

\bibitem[{{Hildebrandt} {et~al.}(2021){Hildebrandt}, {van den Busch}, {Wright}, {Blake}, {Joachimi}, {Kuijken}, {Tr{\"o}ster}, {Asgari}, {Bilicki}, {de Jong}, {Dvornik}, {Erben}, {Getman}, {Giblin}, {Heymans}, {Kannawadi}, {Lin}, \& {Shan}}]{kids1000_photoz}
{Hildebrandt}, H., {van den Busch}, J.~L., {Wright}, A.~H., {et~al.} 2021, \aap, 647, A124, \dodoi{10.1051/0004-6361/202039018}

\bibitem[{{Hsieh} \& {Yee}(2014)}]{dempz}
{Hsieh}, B.~C., \& {Yee}, H.~K.~C. 2014, \apj, 792, 102, \dodoi{10.1088/0004-637X/792/2/102}

\bibitem[{{Ivezi{\'c}} {et~al.}(2019){Ivezi{\'c}}, {Kahn}, {Tyson}, {Abel}, {Acosta}, {Allsman}, {Alonso}, {AlSayyad}, {Anderson}, {Andrew}, {Angel}, {Angeli}, {Ansari}, {Antilogus}, {Araujo}, {Armstrong}, {Arndt}, {Astier}, {Aubourg}, {Auza}, {Axelrod}, {Bard}, {Barr}, {Barrau}, {Bartlett}, {Bauer}, {Bauman}, {Baumont}, {Bechtol}, {Bechtol}, {Becker}, {Becla}, {Beldica}, {Bellavia}, {Bianco}, {Biswas}, {Blanc}, {Blazek}, {Blandford}, {Bloom}, {Bogart}, {Bond}, {Booth}, {Borgland}, {Borne}, {Bosch}, {Boutigny}, {Brackett}, {Bradshaw}, {Brandt}, {Brown}, {Bullock}, {Burchat}, {Burke}, {Cagnoli}, {Calabrese}, {Callahan}, {Callen}, {Carlin}, {Carlson}, {Chandrasekharan}, {Charles-Emerson}, {Chesley}, {Cheu}, {Chiang}, {Chiang}, {Chirino}, {Chow}, {Ciardi}, {Claver}, {Cohen-Tanugi}, {Cockrum}, {Coles}, {Connolly}, {Cook}, {Cooray}, {Covey}, {Cribbs}, {Cui}, {Cutri}, {Daly}, {Daniel}, {Daruich}, {Daubard}, {Daues}, {Dawson}, {Delgado}, {Dellapenna}, {de Peyster}, {de Val-Borro}, {Digel}, {Doherty}, {Dubois},
  {Dubois-Felsmann}, {Durech}, {Economou}, {Eifler}, {Eracleous}, {Emmons}, {Fausti Neto}, {Ferguson}, {Figueroa}, {Fisher-Levine}, {Focke}, {Foss}, {Frank}, {Freemon}, {Gangler}, {Gawiser}, {Geary}, {Gee}, {Geha}, {Gessner}, {Gibson}, {Gilmore}, {Glanzman}, {Glick}, {Goldina}, {Goldstein}, {Goodenow}, {Graham}, {Gressler}, {Gris}, {Guy}, {Guyonnet}, {Haller}, {Harris}, {Hascall}, {Haupt}, {Hernandez}, {Herrmann}, {Hileman}, {Hoblitt}, {Hodgson}, {Hogan}, {Howard}, {Huang}, {Huffer}, {Ingraham}, {Innes}, {Jacoby}, {Jain}, {Jammes}, {Jee}, {Jenness}, {Jernigan}, {Jevremovi{\'c}}, {Johns}, {Johnson}, {Johnson}, {Jones}, {Juramy-Gilles}, {Juri{\'c}}, {Kalirai}, {Kallivayalil}, {Kalmbach}, {Kantor}, {Karst}, {Kasliwal}, {Kelly}, {Kessler}, {Kinnison}, {Kirkby}, {Knox}, {Kotov}, {Krabbendam}, {Krughoff}, {Kub{\'a}nek}, {Kuczewski}, {Kulkarni}, {Ku}, {Kurita}, {Lage}, {Lambert}, {Lange}, {Langton}, {Le Guillou}, {Levine}, {Liang}, {Lim}, {Lintott}, {Long}, {Lopez}, {Lotz}, {Lupton}, {Lust}, {MacArthur}, {Mahabal},
  {Mandelbaum}, {Markiewicz}, {Marsh}, {Marshall}, {Marshall}, {May}, {McKercher}, {McQueen}, {Meyers}, {Migliore}, {Miller}, {Mills}, {Miraval}, {Moeyens}, {Moolekamp}, {Monet}, {Moniez}, {Monkewitz}, {Montgomery}, {Morrison}, {Mueller}, {Muller}, {Mu{\~n}oz Arancibia}, {Neill}, {Newbry}, {Nief}, {Nomerotski}, {Nordby}, {O'Connor}, {Oliver}, {Olivier}, {Olsen}, {O'Mullane}, {Ortiz}, {Osier}, {Owen}, {Pain}, {Palecek}, {Parejko}, {Parsons}, {Pease}, {Peterson}, {Peterson}, {Petravick}, {Libby Petrick}, {Petry}, {Pierfederici}, {Pietrowicz}, {Pike}, {Pinto}, {Plante}, {Plate}, {Plutchak}, {Price}, {Prouza}, {Radeka}, {Rajagopal}, {Rasmussen}, {Regnault}, {Reil}, {Reiss}, {Reuter}, {Ridgway}, {Riot}, {Ritz}, {Robinson}, {Roby}, {Roodman}, {Rosing}, {Roucelle}, {Rumore}, {Russo}, {Saha}, {Sassolas}, {Schalk}, {Schellart}, {Schindler}, {Schmidt}, {Schneider}, {Schneider}, {Schoening}, {Schumacher}, {Schwamb}, {Sebag}, {Selvy}, {Sembroski}, {Seppala}, {Serio}, {Serrano}, {Shaw}, {Shipsey}, {Sick}, {Silvestri},
  {Slater}, {Smith}, {Smith}, {Sobhani}, {Soldahl}, {Storrie-Lombardi}, {Stover}, {Strauss}, {Street}, {Stubbs}, {Sullivan}, {Sweeney}, {Swinbank}, {Szalay}, {Takacs}, {Tether}, {Thaler}, {Thayer}, {Thomas}, {Thornton}, {Thukral}, {Tice}, {Trilling}, {Turri}, {Van Berg}, {Vanden Berk}, {Vetter}, {Virieux}, {Vucina}, {Wahl}, {Walkowicz}, {Walsh}, {Walter}, {Wang}, {Wang}, {Warner}, {Wiecha}, {Willman}, {Winters}, {Wittman}, {Wolff}, {Wood-Vasey}, {Wu}, {Xin}, {Yoachim}, \& {Zhan}}]{lsst}
{Ivezi{\'c}}, {\v{Z}}., {Kahn}, S.~M., {Tyson}, J.~A., {et~al.} 2019, \apj, 873, 111, \dodoi{10.3847/1538-4357/ab042c}

\bibitem[{Izbicki \& Lee(2017)}]{fzboost2017}
Izbicki, R., \& Lee, A.~B. 2017, Electronic Journal of Statistics, 11, 2800 , \dodoi{10.1214/17-EJS1302}

\bibitem[{Jones {et~al.}(2024)Jones, Do, Boscoe, Singal, Wan, \& Nguyen}]{jones2023}
Jones, E., Do, T., Boscoe, B., {et~al.} 2024, The Astrophysical Journal, 964, 130, \dodoi{10.3847/1538-4357/ad2070}

\bibitem[{{Korytov} {et~al.}(2019){Korytov}, {Hearin}, {Kovacs}, {Larsen}, {Rangel}, {Hollowed}, {Benson}, {Heitmann}, {Mao}, {Bahmanyar}, {Chang}, {Campbell}, {DeRose}, {Finkel}, {Frontiere}, {Gawiser}, {Habib}, {Joachimi}, {Lanusse}, {Li}, {Mandelbaum}, {Morrison}, {Newman}, {Pope}, {Rykoff}, {Simet}, {To}, {Vikraman}, {Wechsler}, {White}, \& {(The LSST Dark Energy Science Collaboration}}]{cosmoDC2}
{Korytov}, D., {Hearin}, A., {Kovacs}, E., {et~al.} 2019, \apjs, 245, 26, \dodoi{10.3847/1538-4365/ab510c}

\bibitem[{{LSST-DESC RAIL developer team} {et~al.}(2023){LSST-DESC RAIL developer team}, Schmidt, Gschwend, Crenshaw, Yan, Charles, Malz, Joudaki, Lynn, Tortorelli, hangqianjun, joezuntz, imoskowitz, Kalmbach, jlvdb, sylvielsstfr, Cohen-Tanugi, Santiago, Oldag, DeLucchi, sjs86, vladislav doster, Lanusse, \& Kelly}]{rail}
{LSST-DESC RAIL developer team}, Schmidt, S., Gschwend, J., {et~al.} 2023, LSSTDESC/RAIL: v0.98.5, v0.98.5,  Zenodo, \dodoi{10.5281/zenodo.7927358}

\bibitem[{{Masters} {et~al.}(2015){Masters}, {Capak}, {Stern}, {Ilbert}, {Salvato}, {Schmidt}, {Longo}, {Rhodes}, {Paltani}, {Mobasher}, {Hoekstra}, {Hildebrandt}, {Coupon}, {Steinhardt}, {Speagle}, {Faisst}, {Kalinich}, {Brodwin}, {Brescia}, \& {Cavuoti}}]{som_followup}
{Masters}, D., {Capak}, P., {Stern}, D., {et~al.} 2015, \apj, 813, 53, \dodoi{10.1088/0004-637X/813/1/53}

\bibitem[{Miyatake {et~al.}(2023)Miyatake, Sugiyama, Takada, Nishimichi, Li, Shirasaki, More, Kobayashi, Nishizawa, Rau, Zhang, Takahashi, Dalal, Mandelbaum, Strauss, Hamana, Oguri, Osato, Luo, Kannawadi, Hsieh, Armstrong, Bosch, Komiyama, Lupton, Lust, MacArthur, Miyazaki, Murayama, Okura, Price, Sunayama, Tait, Tanaka, \& Wang}]{hsc_y3_results2}
Miyatake, H., Sugiyama, S., Takada, M., {et~al.} 2023, Phys. Rev. D, 108, 123517, \dodoi{10.1103/PhysRevD.108.123517}

\bibitem[{Moskowitz {et~al.}(2023)Moskowitz, Gawiser, Bault, Broussard, Newman, \& Zuntz}]{Moskowitz_2023}
Moskowitz, I., Gawiser, E., Bault, A., {et~al.} 2023, The Astrophysical Journal, 950, 49, \dodoi{10.3847/1538-4357/accc88}

\bibitem[{Myles {et~al.}(2021)Myles, Alarcon, Amon, Sánchez, Everett, DeRose, McCullough, Gruen, Bernstein, Troxel, Dodelson, Campos, MacCrann, Yin, Raveri, Amara, Becker, Choi, Cordero, Eckert, Gatti, Giannini, Gschwend, Gruendl, Harrison, Hartley, Huff, Kuropatkin, Lin, Masters, Miquel, Prat, Roodman, Rykoff, Sevilla-Noarbe, Sheldon, Wechsler, Yanny, Abbott, Aguena, Allam, Annis, Bacon, Bertin, Bhargava, Bridle, Brooks, Burke, Carnero Rosell, Carrasco Kind, Carretero, Castander, Conselice, Costanzi, Crocce, da Costa, Pereira, Desai, Diehl, Eifler, Elvin-Poole, Evrard, Ferrero, Ferté, Flaugher, Fosalba, Frieman, García-Bellido, Gaztanaga, Giannantonio, Hinton, Hollowood, Honscheid, Hoyle, Huterer, James, Krause, Kuehn, Lahav, Lima, Maia, Marshall, Martini, Melchior, Menanteau, Mohr, Morgan, Muir, Ogando, Palmese, Paz-Chinchón, Plazas, Rodriguez-Monroy, Samuroff, Sanchez, Scarpine, Secco, Serrano, Smith, Soares-Santos, Suchyta, Swanson, Tarle, Thomas, To, Varga, Weller, \& Wester}]{des_y3_photoz}
Myles, J., Alarcon, A., Amon, A., {et~al.} 2021, Monthly Notices of the Royal Astronomical Society, 505, 4249, \dodoi{10.1093/mnras/stab1515}

\bibitem[{Rau {et~al.}(2023)Rau, Dalal, Zhang, Li, Nishizawa, More, Mandelbaum, Miyatake, Strauss, \& Takada}]{hsc_photoz}
Rau, M.~M., Dalal, R., Zhang, T., {et~al.} 2023, Monthly Notices of the Royal Astronomical Society, 524, 5109, \dodoi{10.1093/mnras/stad1962}

\bibitem[{{Schmidt} {et~al.}(2020){Schmidt}, {Malz}, {Soo}, {Almosallam}, {Brescia}, {Cavuoti}, {Cohen-Tanugi}, {Connolly}, {DeRose}, {Freeman}, {Graham}, {Iyer}, {Jarvis}, {Kalmbach}, {Kovacs}, {Lee}, {Longo}, {Morrison}, {Newman}, {Nourbakhsh}, {Nuss}, {Pospisil}, {Tranin}, {Wechsler}, {Zhou}, {Izbicki}, \& {LSST Dark Energy Science Collaboration}}]{schmidt2020}
{Schmidt}, S.~J., {Malz}, A.~I., {Soo}, J.~Y.~H., {et~al.} 2020, \mnras, 499, 1587, \dodoi{10.1093/mnras/staa2799}

\bibitem[{Shorten \& Khoshgoftaar(2019)}]{augmentation}
Shorten, C., \& Khoshgoftaar, T. 2019, Journal of Big Data, 6, 60, \dodoi{10.1186/s40537-019-0197-0}

\bibitem[{Springel(2005)}]{gadget2}
Springel, V. 2005, Monthly Notices of the Royal Astronomical Society, 364, 1105–1134, \dodoi{10.1111/j.1365-2966.2005.09655.x}

\bibitem[{Stylianou {et~al.}(2022)Stylianou, Malz, Hatfield, Crenshaw, \& Gschwend}]{Stylianou2022}
Stylianou, N., Malz, A.~I., Hatfield, P., Crenshaw, J.~F., \& Gschwend, J. 2022, Publications of the Astronomical Society of the Pacific, 134, 044501, \dodoi{10.1088/1538-3873/ac59bf}

\bibitem[{Sugiyama {et~al.}(2023)Sugiyama, Miyatake, More, Li, Shirasaki, Takada, Kobayashi, Takahashi, Nishimichi, Nishizawa, Rau, Zhang, Dalal, Mandelbaum, Strauss, Hamana, Oguri, Osato, Kannawadi, Hsieh, Luo, Armstrong, Bosch, Komiyama, Lupton, Lust, Miyazaki, Murayama, Okura, Price, Tait, Tanaka, \& Wang}]{hsc_y3_results1}
Sugiyama, S., Miyatake, H., More, S., {et~al.} 2023, Phys. Rev. D, 108, 123521, \dodoi{10.1103/PhysRevD.108.123521}

\bibitem[{van~den Busch {et~al.}(2022)van~den Busch, Wright, Hildebrandt, Bilicki, Asgari, Joudaki, Blake, Heymans, Kannawadi, Shan, \& Tröster}]{kids_som}
van~den Busch, J.~L., Wright, A.~H., Hildebrandt, H., {et~al.} 2022, Astronomy \& Astrophysics, 664, A170, \dodoi{10.1051/0004-6361/202142083}

\bibitem[{Weaver {et~al.}(2022)Weaver, Kauffmann, Ilbert, McCracken, Moneti, Toft, Brammer, Shuntov, Davidzon, Hsieh, Laigle, Anastasiou, Jespersen, Vinther, Capak, Casey, McPartland, Milvang-Jensen, Mobasher, Sanders, Zalesky, Arnouts, Aussel, Dunlop, Faisst, Franx, Furtak, Fynbo, Gould, Greve, Gwyn, Kartaltepe, Kashino, Koekemoer, Kokorev, Le~Fèvre, Lilly, Masters, Magdis, Mehta, Peng, Riechers, Salvato, Sawicki, Scarlata, Scoville, Shirley, Silverman, Sneppen, Smolc̆ić, Steinhardt, Stern, Tanaka, Taniguchi, Teplitz, Vaccari, Wang, \& Zamorani}]{cosmos2020}
Weaver, J.~R., Kauffmann, O.~B., Ilbert, O., {et~al.} 2022, The Astrophysical Journal Supplement Series, 258, 11, \dodoi{10.3847/1538-4365/ac3078}

\bibitem[{Wechsler {et~al.}(2022)Wechsler, DeRose, Busha, Becker, Rykoff, \& Evrard}]{add_gals}
Wechsler, R.~H., DeRose, J., Busha, M.~T., {et~al.} 2022, The Astrophysical Journal, 931, 145, \dodoi{10.3847/1538-4357/ac5b0a}

\bibitem[{Zuntz {et~al.}(2021)Zuntz, Lanusse, Malz, Wright, Slosar, Abolfathi, Alonso, Bault, Bom, Brescia, Broussard, Campagne, Cavuoti, Cypriano, Fraga, Gawiser, Gonzalez, Green, Hatfield, Iyer, Kirkby, Nicola, Nourbakhsh, Park, Teixeira, Heitmann, Kovacs, \& and}]{tomo_challenge}
Zuntz, J., Lanusse, F., Malz, A.~I., {et~al.} 2021, The Open Journal of Astrophysics, 4, \dodoi{10.21105/astro.2108.13418}

\end{thebibliography}

\end{document}